\documentclass[aps, preprint, one column, showkeys,showpacs,amsmath,amssymb,superscriptaddress]{revtex4-1}

\usepackage{lineno,hyperref}
\usepackage{todonotes}
\usepackage{makeidx}
\usepackage{graphics}
\usepackage{color}
\usepackage{listings}
\usepackage{bm}
\usepackage{url}
\usepackage{booktabs}
\usepackage{amsmath} 
\usepackage{paralist}
\usepackage{amssymb}
\usepackage{amsthm}
\usepackage{subcaption} 
\usepackage{algorithm}
\usepackage{algorithmic}
\usepackage{lipsum}
\usepackage{combelow}
\usepackage{ulem}
\usepackage[percent]{overpic}

\newcommand*\diff{\mathop{}\!\mathrm{d}}

\begin{document}

 
\title{Fast kinetic simulator for relativistic matter}
\author{V. E. Ambru\cb{s}}
\affiliation{Institut f\"ur Theoretische Physik, Johann Wolfgang 
Goethe-Universit\"at, Max-von-Laue-Strasse 1, D-60438 Frankfurt am Main, Germany}
\affiliation{Department of Physics, West University of Timi\cb{s}oara, 
Bd.~Vasile P\^arvan 4, Timi\cb{s}oara 300223, Romania}
\author{L. Bazzanini}
\affiliation{Universit\`a di Ferrara and INFN-Ferrara, I-44122 Ferrara,~Italy}
\author{A. Gabbana}
\affiliation{Eindhoven University of Technology, 5600 MB Eindhoven,~ The 
Netherlands}
\author{D. Simeoni}
\affiliation{Universit\`a di Ferrara and INFN-Ferrara, I-44122 Ferrara,~Italy}
\affiliation{Bergische Universit\"at Wuppertal, D-42119 Wuppertal,~Germany}
\affiliation{University of Cyprus, Physics department, CY-1678 Nicosia,~Cyprus}
\author{S. Succi}
\affiliation{Center for Life Nano Science @ La Sapienza, Italian Institute of 
Technology, Viale Regina Elena 295, I-00161 Roma,~Italy}
\affiliation{Istituto Applicazioni del Calcolo, National Research Council of 
Italy, Via dei Taurini 19, I-00185 Roma,~Italy}
\author{R. Tripiccione}
\affiliation{Universit\`a di Ferrara and INFN-Ferrara, I-44122 Ferrara,~Italy}

\begin{abstract}
We present a new family of relativistic lattice kinetic schemes for the 
efficient simulation of relativistic flows in both strongly-interacting (fluid)
and weakly-interacting (rarefied gas) regimes. 
The method can also deal with both massless and massive particles, thereby 
encompassing ultra-relativistic and mildly-relativistic regimes alike.
The computational performance of the method for the
simulation of relativistic flows across the aforementioned 
regimes is discussed in detail, along with prospects of future
applications for Quark-Gluon Plasma, electron flows in 
graphene and systems in astrophysical contexts. 
\end{abstract}

\maketitle

\section{Introduction}

Relativistic fluid dynamics deals with the study of the motion of 
particles traveling close to the speed of light,  as it is typically the 
case in plasma physics, astrophysics and cosmology~\cite{rezzolla-book-2013}.
In recent years, experimental data from high-energy particles colliders, such 
as RHIC and LHC, have provided clearcut evidence 
that the exotic state of matter known as Quark Gluon Plasma (QGP) 
also behaves like a low-viscosity relativistic fluid ~\cite{florkowski-rpp-2018}.
Furthermore, additional evidence started to emerge that electron fluids 
in exotic two-dimensional materials, like for example graphene, are also 
described by relativistic hydrodynamics~\cite{lucas-jopcm-2018}.
More generally, in light of the AdS-CFT duality~\cite{maldacena-ijtp-1999},
relativistic hydrodynamics has acquired a very distinct role as a low-energy
effective field theory at the crossroad between high-energy physics, gravity 
and quantum condensed matter
~\cite{romatschke-book-2019,lucas-pnas-2016,succi-epl-2015}. 

The physics of fluids, classical, quantum and relativistic alike, is
characterized by the subtle competition between mechanisms which
promote equilibrium (collisions) and mechanisms which sustain
the opposite tendency (transport): 
in Boltzmann momentous words, "the evershifting battle"
between equilibrium and non-equilibrium
~\cite{boltzmann-book-2020}. 

In relativistic fluids the above competition is controlled by two 
dimensionless groups, the \textit{relativistic coldness} $\zeta = mc^2/k_BT$,
ratio of the particle rest energy to the thermal energy, 
and the \textit{Knudsen number} $\rm{Kn} = \lambda/\ell$.
Here $m$ is the mass of the particle, $c$ the speed of light, $T$ the temperature,
$k_B$ the Boltzmann's constant, $\lambda$ the mean free path and $\ell$ is a 
characteristic macro scale. 

The relativistic coldness scales like the inverse temperature, hence 
it takes large values in the non-relativistic regime where 
kinetic energy is small as compared to the rest energy. 
It also scales linearly with the particle mass, which means that
high values of the coldness correspond to heavy particles, pointing
again to the non-relativistic regime.
Importantly, the relativistic coldness is an equilibrium property. 

The Knudsen number, on the other hand, measures the departure 
from (local) equilibrium due to the spatial inhomogeneities 
that drive transport phenomena and dissipation.
Since the mean free path scales like the inverse density, so does
the Knudsen number, which takes up substantial
values in the rarefied gas regime, where the hydrodynamic 
description no longer holds.

In broad strokes the $(\zeta-\rm{Kn})$ plane can be split in 
four quadrants:

\begin{itemize}
  \item[1)] Relativistic fluids     ($\zeta<1, \rm{Kn}<0.01$); 
  \item[2)] Non-relativistic fluids ($\zeta>1, \rm{Kn}<0.01$);
  \item[3)] Relativistic gases      ($\zeta<1, \rm{Kn}>0.01$); 
  \item[4)] Non-relativistic gases  ($\zeta>1, \rm{Kn}>0.01$); 
\end{itemize}

The four quadrants above encompass a broad class of vastly
different states of matter, from QGP (1), to 
Bose-Einstein condensates (2), to relativistic and classical 
astrophysical systems (3 and 4).  

Clearly, no single numerical method can work seamlessly across  
such broad variety of systems, a typical separation 
being between fluid-dynamic methods based on the discretisation 
of the fluid equations~\cite{Rischke:1995ir,huovinen-plb-2001,aguiar-jpg-2000,schenke-prc-2010, 
molnar-epjc-2010,Gerhard:2012uf,delzanna-epjc-2013,Karpenko:2013wva,Pandya:2022pif}
 and kinetic methods (mostly Monte Carlo) for the Boltzmann equation
~\cite{nonaka-epjc-2000,xu-prc-2007,Petersen:2008dd,plumari-prc-2012,Weil:2016zrk,Gallmeister:2018mcn}.

Even though lattice kinetic methods offer a potential bridge between these 
two main families, to date, they have been confined 
to the relativistic fluid sector only ~\cite{mendoza-prl-2010,mendoza-prd-2010,gabbana-pr-2020}, 
and successfully applied to a number of relativistic hydrodynamic problems
in QGP \cite{romatschke-prc-2011,romatschke-prd-2012}, electron transport in 
graphene~\cite{gabbana-cf-2018} 
and also cosmic neutrino transport~\cite{weih-mnras-2020}.
A similar approach has been directed to the study of   
ultra-relativistic gases in $(3+1)$~\cite{ambrus-prc-2018}, as well as in 
$(2+1)$ dimensions~\cite{coelho-cf-2018,bazzanini-jocs-2021}.

In this paper we extend the lattice kinetic approach to higher order 
discrete velocity sets that allow to handle finite values of the Knudsen number. 
This approach applies to both massive and massless particles, thereby extending
the range of applicability of the method along both directions in the 
$(\zeta-\rm{Kn})$ parameter plane.
As a result, the present method is expected to offer a useful complement to 
current QGP codes, such as vSHASTA \cite{Molnar:2008fv}, MUSIC~\cite{schenke-prc-2010} 
or vHLLE \cite{Karpenko:2013wva}, in assisting 
the experimental activity of the existing collaborations, such as 
PHENIX~\cite{adler-prc-2004}, PHOBOS~\cite{adams-prc-2005}, 
BRAHMS~\cite{bearden-prl-2005}, STAR~\cite{back-prc-2005} at RHIC, and 
ALICE~\cite{adam-prl-2016}, ATLAS~\cite{aad-prl-2015}, and 
CMS~\cite{sirunyan-prl-2020} at LHC.

\section{Results}\label{sec:numerics}

\subsection{Model Overview}\label{subsec:overview}

In this work we introduce an extension of a numerical method, the Relativistic 
Lattice Boltzmann Method, originally designed for the study or relativistic 
fluids, which is capable of accurately solving the relativistic Boltzmann 
equation in the Relaxation-Time Approximation (RTA) for a broad set of kinematic
regimes.

The key insight in the development of Lattice Boltzmann Methods is the 
realization that the Boltzmann transport equation, (in this case expressed in 
the language of special relativity), can appropriately be truncated and 
discretized to recover the dynamics at the hydro level. This operation leads to
an evolution equation for the probability density function of particle position
and momentum, whose moments deliver the sought after expressions for the 
hydrodynamic fields.
In particular, the key ingredient to the simulation of weakly interacting regimes
is represented by a controlled discretization of the momentum space, which is
based on the product of two high-order quadrature rules that discretize separately
the various components of momentum: a Gauss-Laguerre rule of order $N$ is employed
for the energy component, and quadrature rules of order $K$ for the integration of
functions on the sphere ~\cite{ahrens-prsA-2009} are considered for the
remaining momentum components. The orders $N$ and $K$ of the quadratures employed
lead to a number of $N_{\rm pop}$ discrete momenta. 
The reader is refereed to Sec.~\ref{sec:model-description} for full details on
the numerical methods, while in this section we focus on a few examples of 
applications and benchmarks which highlight the enhanced accuracy of the present
scheme in rarefied conditions.

\subsection{Shock Waves in Quark Gluon Plasma}\label{subsec:monoshock}

\begin{figure*}
  \includegraphics[width=0.99\textwidth]{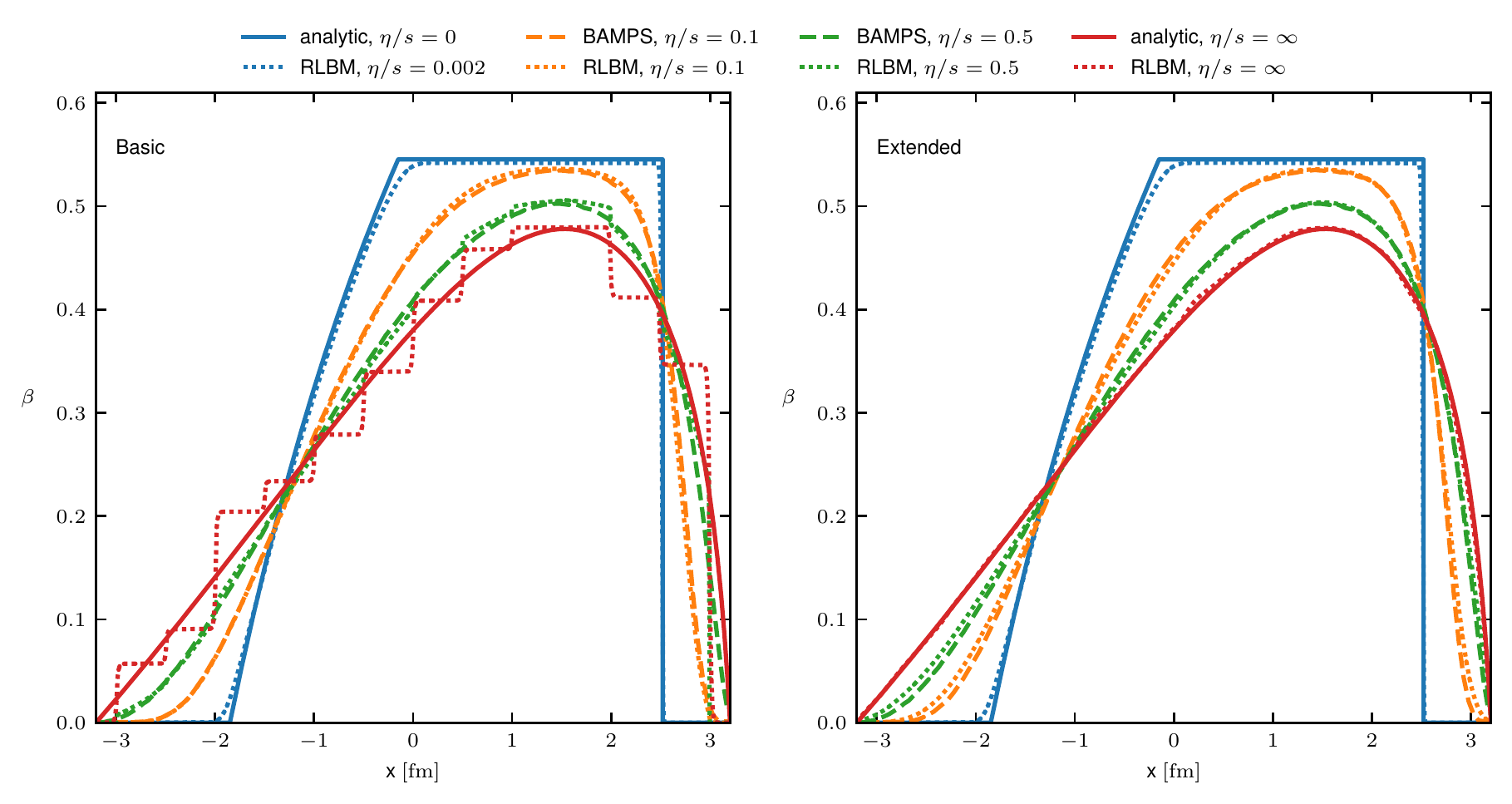}
  \caption{ Riemann problem for an ultra-relativistic gas of particles for various viscous regimes. 
            We show the macroscopic velocity profile $\beta = U^x / U^0$ at $t = 3.2~\rm{fm}/c$,
            comparing the results of a RLBM model developed for the hydrodynamic regime~\cite{gabbana-prc-2020} (left), 
            and the high order scheme described in this work (right), which 
            allows to significantly increase the accuracy of the numerical results
            in beyond-hydrodynamic regimes ($\eta /s > 0.1$).
            }\label{fig:sod1d_comparison}
\end{figure*}

\begin{figure*}[htb]
  \includegraphics[width=0.99\textwidth]{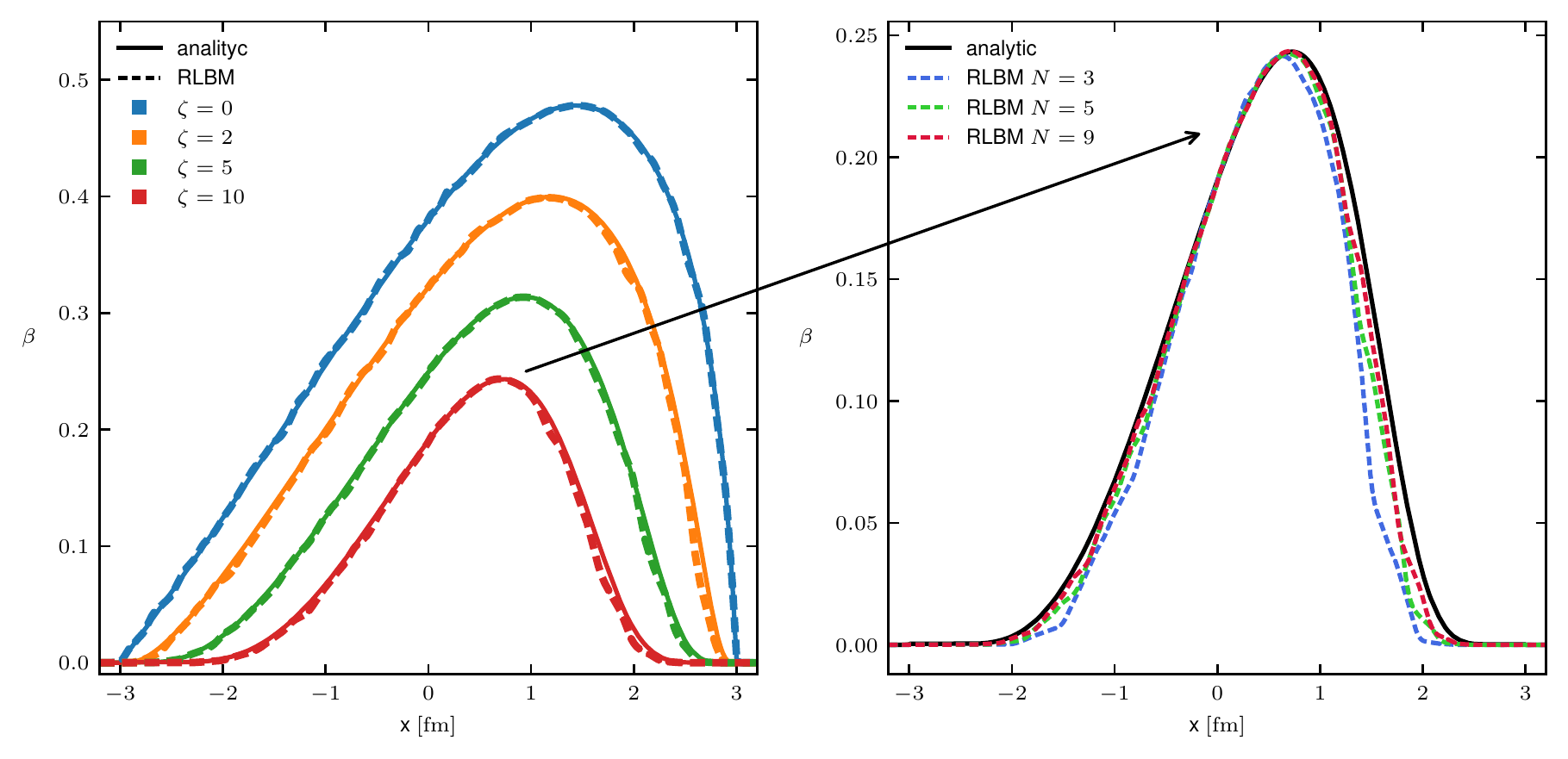}
  \caption{ Riemann problem for a relativistic gas of particles in the free-streaming regimes
            at different values of the rest mass $m = (0, 0.8, 2, 4)$ GeV, corresponding to 
            $\zeta = (0, 2, 5, 10)$. 
            The left panel shows the macroscopic velocity profile $\beta = U^x / U^0$ at $t = 3~\rm{fm}/c$,
            comparing the results of RLBM against analytic solutions. On the left panel, 
            we use $N=3$, $K=19$ for $\zeta = (0, 2)$, and $N=9$, $K=19$ for $\zeta = (5, 10)$.
            The right panel shows the effect of increasing the radial quadrature for the
            case $\zeta = 10$.
           }
  \label{sod1d_diffmass}
\end{figure*}

We start our numerical analysis by considering the relativistic Riemann problem.
This problem describes a tube filled with a gas which initially is 
in two different states (particle number density $n$, temperature $T$ and macroscopic velocity $U^\alpha$) on the two sides of a membrane placed at $x = 0$:
\begin{align}\label{eq:1dsod_initial_conditions}
  \left(n, T, U^x \right) = 
  \begin{cases}
  n_L, T_L, 0  \quad x<0 \\
  n_R, T_R, 0  \quad x>0 \\
  \end{cases} \quad . 
\end{align}
Once the membrane is removed, the system develops one-dimensional shock/rarefaction 
waves traveling along the $x-$axis. 
We use the same initial conditions as in~\cite{gabbana-prc-2020}, namely
\begin{equation}\label{eq:1dsod_init}
  \begin{cases}
    n_{\mathrm{L}}=13.575\,\mathrm{fm}^{-3}, & T_{\mathrm{L}}=400\,\mathrm{MeV}\\
    n_{\mathrm{R}}=1.65  \,\mathrm{fm}^{-3}, & T_{\mathrm{R}}=200\,\mathrm{MeV}
  \end{cases} \quad.
\end{equation}
In these simulations we keep the ratio between the shear viscosity and the 
entropy density, $\eta / s$, fixed to a constant value.
For the parameter $\eta$ we use the analytic expressions resulting from the first-order
Chapman-Enskog expansion~\cite{gabbana-pr-2020},
while the entropy density is approximated using~\cite{cercignani-book-2002}
\begin{equation}\label{eq:entrophy-density}
  s = n \left( \zeta \frac{K_3(\zeta)}{K_2(\zeta)} - \ln{\left(\frac{n}{n^{\rm eq}}\right)} \right) \quad ,
\end{equation}
with the equilibrium density given by
\begin{equation}\label{eq:equilibrium-density}
  n^{\rm eq} = d_{\rm G} \frac{T^3}{2 \pi^2} \zeta^2 K_2(\zeta) \quad ,
\end{equation}
$d_{\rm G} = 16$ being the degeneracy factor of gluons and $K_\nu(\zeta)$ the 
Modified Bessel function of second kind.

In the left panel of Fig.~\ref{fig:sod1d_comparison} we have reproduced, for reference,
the results presented in Fig.1 from~\cite{gabbana-prc-2020}, showing 
the profile of the macroscopic velocity $\beta = U^x / U^0$ for a gas of massless particles
in several different kinematic regimes. The two limiting cases, corresponding to the 
inviscid ($\eta /s \rightarrow 0$) and the ballistic ($\eta /s \rightarrow \infty$) regimes
admit an analytic solution, whereas for intermediate regimes we compare the results
against BAMPS (Boltzmann Approach to Multi-Parton Scatterings)~\cite{xu-prc-2007}, 
which solves the Boltzmann equation using a Monte Carlo technique.
The on-lattice RLBM model correctly reproduces the solution both in the inviscid 
and in the hydrodynamic regime ($\eta /s = 0.1$). On the other hand, for larger values 
of $\eta /s$, as we move beyond the hydrodynamic regime, the macroscopic velocity profile
develops artifacts which become most apparent in the ballistic limit.
In the right panel of Fig.~\ref{fig:sod1d_comparison} we show that by employing high order
off-lattice quadratures~\cite{ambrus-prc-2018} it is possible to improve the accuracy 
even in rarefied conditions. Here we have used a radial quadrature of order $N = 3$ and 
an angular quadrature of order $K = 15$ with $N_K = 120$ discrete components 
(more details on the meaning of these values in Sec.~\ref{sec:model-description}).
We remark that the off-lattice model preserves the same level of accuracy of the on-lattice
scheme also in the hydrodynamic regimes.

We now turn to the analysis of a relativistic gas of massive particles;
we consider once again the initial conditions given in Eq.~\ref{eq:1dsod_init},
with $m = (0, 0.8, 2, 4)$ GeV. 
In the simulations we normalize quantities with respect to $T_L$ and $n_L$ 
(see Eq.~\eqref{eq:1dsod_initial_conditions}), corresponding 
to $\zeta = (0, 2, 5, 10)$.

In the left panel of Fig.~\ref{sod1d_diffmass} we compare the results obtained 
in the free-streaming regime, corresponding to $\eta / s \rightarrow \infty$, 
against analytic solutions finding again a satisfactory 
match between the two.

One important remark is that, as the rest mass of the gas increases, we need
to employ a higher order radial quadrature to match the same 
level of accuracy achieved, for example, in the massless case. 
This is shown in the right panel of Fig.~\ref{sod1d_diffmass}, where we compare the 
results at $\zeta = 10$ obtained by keeping fixed the angular quadrature at
$K = 19$ and varying the radial quadrature from $N = 3$ (which is the value 
used for the massless case)
up to $N = 9$. The results show the improvements achieved by increasing the degree
of accuracy of the radial quadrature. 

We now further investigate how the accuracy of the method depends on 
the degree of angular and radial quadrature, studying the Riemann problem
at different values of Knudsen number and relativistic coldness.

We use the same initial conditions given by Eq.~\eqref{eq:1dsod_init}.
In order to assess the rarefied regime, we make use of 
the numerical Knudsen number, defined as follows:
\begin{align}\label{eq:knudsen}
  \mathrm{Kn} = \frac{\tau \langle v \rangle}{L}   \quad ,
\end{align}
where $L$ defines the spatial resolution chosen for the grid, $\tau$ is the 
relaxation time, and $\langle v \rangle$ is a relative mean velocity, of order 
$1$ in lattice units.
We consider a grid of $L = 1600$ points representing a physical domain of $6.4~\rm{fm}$.
As a reference, we take the results of RLBM simulations with a high
resolution in terms of both grid and momentum discretization,
namely $L=6400$, $N=9$ and $K = 31$.

We define the L2-relative error with respect to the temperature field as
\begin{align}\label{eq:relative_error}
  \epsilon = \frac{|| T - T_{\rm hr}||_2}{||T_{\rm hr}||_2} \quad ,
\end{align}
where ``${\rm hr}$'' refers to high resolution simulations.

\begin{figure}
  \centering
  \includegraphics[width=0.79\columnwidth]{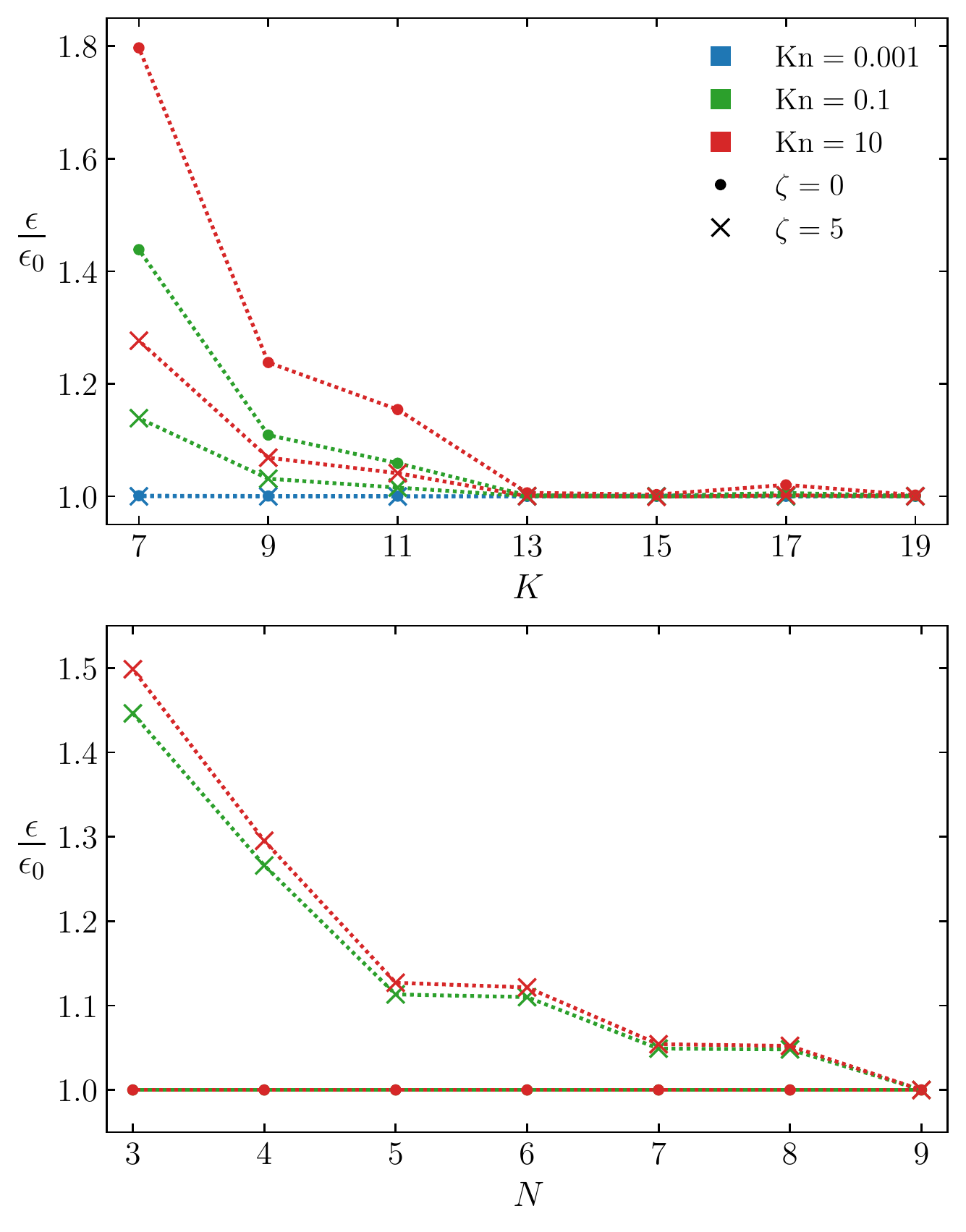}
  \caption{ Discretization error versus the radial/angular quadrature 
            at different Knudsen numbers and for different values of the relativistic coldness.
            The relative error $\epsilon$ (Eq.~\ref{eq:relative_error}) 
            is normalized with respect to its asymptotic value $\epsilon_0$, defined as $\epsilon_0=\epsilon(K=19)$ 
            in the top panel and $\epsilon_0=\epsilon(N=9)$ in the bottom panel.
            Top: Effect of increasing the degree of the angular quadrature, while keeping the 
                 radial quadrature fixed at $N = 3$.
            Bottom: Effect of increasing the degree of the radial quadrature, while keeping the 
                 angular quadrature fixed at $K = 13$.
          } \label{fig:quad-conv} 
\end{figure}

In the top panel of Fig.~\ref{fig:quad-conv}, we plot this observable versus 
the order of the angular quadrature $K$. 
All simulations run at fixed order of the radial 
quadrature, $N=3$, and different values of relativistic coldness, namely $\zeta=0$ and $\zeta=5$.

We can observe the effect of increasing 
the order of the angular quadrature $K$, by keeping fixed the radial quadrature to $N=3$. 
We notice that, as the Knudsen number is increased, a higher order leads to significant gains; 
interestingly this effect is more pronounced in the massless case. We also observe that
in the hydrodynamic regime the accuracy is not affected by the quadrature degree.

In the bottom panel of Fig.~\ref{fig:quad-conv} we consider instead the effect of
varying the degree of the radial quadrature $N$, while keeping the angular 
quadrature fixed at $K=13$.
We observe that in the massless case the error does not depend on the radial quadrature,
regardless of the Knudsen number employed.
On the other hand, for $\zeta = 5$ it is necessary to
increase the radial quadrature up to $N = 9$ before saturating the error.

The intuition behind these results is as follows: since massless particles
all travel at the speed of light, their discretized components necessarily lie
on the surface of a sphere in momentum space. As a result, increasing the order of the 
angular quadrature offers a better approximation of momentum space.
On the other hand, massive particles cover a finite range of velocities,
thus requiring tuning of both the radial and the angular components
(see Sec~\ref{subsec:discretization} for more details).

\subsection{Bjorken attractor} \label{subsec:attractor}

\begin{figure*}
  \centering
  \includegraphics[width=0.9\linewidth]{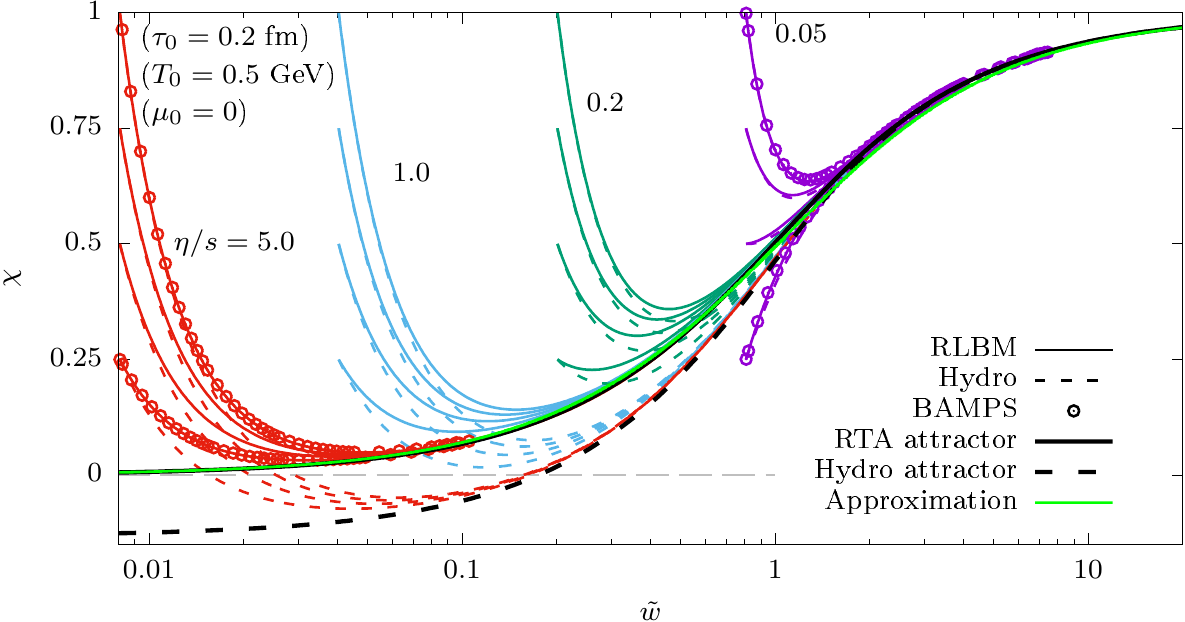} \\
  \caption{
Dependence of $\chi$ \eqref{eq:Bjorken_chi} on $\tilde{w}$ 
\eqref{eq:Bjorken_wtilde} as obtained using RLBM (solid lines), hydro 
(dashed lines) and BAMPS (points). The initial conditions are $\tau_0 = 0.2\ {\rm fm}$ 
and $T_0 = 0.5\ {\rm GeV}$ in all cases, while $\eta / s$ takes the values $5$ 
(red), $1$ (blue), $0.2$ (green) and $0.05$ (purple). The hydro results are 
obtained by directly integrating Eq.~\eqref{eq:Bjorken_eqs}. The BAMPS data are 
taken from Ref.~\cite{Ambrus:2021sjg}. The RLBM algorithm employs $2$ points 
along the radial direction, $20$ points for 
$v_z = \cos \theta$ and $1$ point in the azimuthal direction, as discussed in 
Refs.~\cite{ambrus-prc-2018,ambrus-aip-2019}. The hydro attractor is shown with 
the dashed black line. The RTA attractor obtained using our code (solid black line) 
is in excellent agreement with the analytical approximation from 
Ref.~\cite{Romatschke:2017vte} (solid green line).
}
\label{fig:bjorken}
\end{figure*}

The method described so far is based on a uniform Cartesian grid. 
However, a similar procedure applies
to curvilinear coordinates as well~\cite{ambrus-prc-2018,ambrus-aip-2019},
which represent the most natural choice for a variety of relativistic flow 
problems.
In this subsection we focus on such one case, namely the Bjorken model~\cite{Bjorken:1982qr}, 
which is particularly relevant to QGP experiments. Indeed, the existence of 
Bjorken attractors \cite{Heller:2015dha} and the details of their structure have 
received significant attention in the recent years, for they provide valuable 
information on the initial conditions right after the collisions, the onset of 
fluid-dynamic behavior and also on the material properties of the QGP state of 
matter (see Ref.~\cite{Soloviev:2021lhs} for a recent review). 
The evolution of the system at very early times (the pre-equilibrium phase) 
along the attractor allows one to 
estimate the work done by the QGP plasma against longitudinal expansion, which 
leads to the overall cooling of the fireball \cite{Kurkela:2019kip}. Moreover, 
the entropy production 
in this pre-equilibrium phase can be used to increase the accuracy of the 
connection between the initial state energy and of the overall particle yields 
\cite{Giacalone:2019ldn}.
Another effect of the pre-equilibrium stage is related to the inhomogeneous 
cooling of the initial state, leading to modifications of the transverse profile
that decrease its eccentricity, thus affecting the buildup of flow harmonics, 
such as elliptic flow $v_2$, during the transverse expansion phase 
\cite{Ambrus:2021fej}.

The focus of this section is on the early-time dynamics of the 
fireball induced by the rapid longitudinal expansion, which 
is well described by Bjorken's approximation of longitudinal boost
invariance. For simplicity, we focus on the flow of massless 
particles (with equation of state $\epsilon = 3P$) with 
constant $\eta / s$.

Neglecting the dynamics in the transverse plane, the velocity profile satisfying
 boost-invariance along the $z$ direction is 

\begin{equation}
    U^\alpha \partial_\alpha = \frac{1}{\tau} (t \partial_t + z\partial_z) = \partial_\tau, \label{eq:Bjorken_U}
\end{equation}

where $\tau = \sqrt{t^2 - z^2}$ is the Bjorken time. Employing the Bjorken 
coordinates $(\tau, x, y, \eta_s)$, where $\eta_s = {\rm atanh}(z/t)$
is the pseudorapidity, the particle four flow $N^\mu$ and stress-energy 
tensors $T^{\mu\nu}$ reduce to $N^\tau = n$, $N^x = N^y = N^\eta = 0$, while 
$T^{\mu\nu} = {\rm diag}(\epsilon, P_T , P_T, \tau^{-2} P_L)$.
The diffusion current vanishes, while the pressure deviator $\pi^{\mu\nu}$ takes 
a diagonal form, $\pi^{\mu\nu} = {\rm diag}(0, -\pi_d/2, -\pi_d/2, \tau^{-2} \pi_d)$, 
where $\pi_d = \frac{2}{3}(P_L - P_T)$. Our focus in this section will be on 
the function

\begin{equation}
 \chi = \frac{P_L}{P_T} = \frac{P + \pi_d}{P - \frac{1}{2} \pi_d},
 \label{eq:Bjorken_chi}
\end{equation}

representing the ratio between the longitudinal and transverse pressures.

The macroscopic equations $\nabla_\mu N^\mu = 0$ and 
$\nabla_\mu T^{\mu\nu} = 0$ reduce to

\begin{subequations} \label{eq:Bjorken_eqs}
\begin{align}
 \tau \frac{\partial (\tau n)}{\partial \tau} =& 0,                        \label{eq:Bjorken_eqs_n}\\
 \tau \frac{\partial \epsilon}{\partial \tau} + \epsilon + P + \pi_d =& 0. \label{eq:Bjorken_eqs_e}
\end{align}
The equation for $\pi_d$ can be derived in the frame of the Israel-Stewart 
second order hydrodynamics \cite{israel-anp-1976,israel-pla-1976}, or directly 
from the Boltzmann equation \eqref{eq:boltz_eq} using the Chapman-Enskog 
method \cite{Jaiswal:2013npa},
\begin{equation}
 \tau \frac{\partial \pi_d}{\partial \tau} + \left(\lambda + \frac{\tau}{\tau_R}\right) \pi + \frac{16}{15} P = 0,
 \label{eq:Bjorken_eqs_pi}
\end{equation}
\end{subequations}

where $\lambda = 38/21$ and $\tau_R = 5 \eta / 4 P$ is the Anderson-Witting 
relaxation time. Since $P$ and $\pi_d$ depend only on $\tau$, 
Eq.~\eqref{eq:Bjorken_eqs} can be solved straightforwardly, e.g. using 
Runge-Kutta time stepping. 

The evolution of $\pi_d$ can be obtained from the kinetic equation by writing Eq.~\eqref{eq:boltz_eq} with respect to the Bjorken coordinates, taking into account the degrees of freedom $p = p^\tau$, $v_z = \tau p^\eta / p$ and $\varphi = \arctan(p^y / p^x)$ \cite{Kurkela:2019kip}
\begin{equation}
 \left(\partial_\tau -\frac{v_z(1-v_z^2)}{\tau}\partial_{v_z}-\frac{v_z^2p}{\tau}\partial_{p}\right)f
 = -\frac{v^\mu u_\mu}{\tau_R} (f-f^{\rm eq}),
 \label{eq:boltz_bjorken}
\end{equation}
where $v_z = \cos \theta$ in the language of Sec.~\ref{subsec:discretization}. For simplicity, $f^{\rm eq}$ is
taken as the Maxwell-J\"uttner distribution given in Eq.~\eqref{eq:feq_dist}, which reduces in the case of massless particles to
\begin{equation}
 f^{\rm eq} = \frac{n}{8\pi T^3} \exp\left(-\frac{p^\alpha U_\alpha}{T}\right).
\end{equation}
At initial time, the distribution function is set to the Romatschke-Strickland 
distribution \cite{Romatschke:2003ms,Florkowski:2013lya},
\begin{equation}
 f_{\rm RS} = \frac{g e^{\alpha_0}}{(2\pi)^3} 
 \exp \left[-\frac{1}{\Lambda_0} \sqrt{(p \cdot u)^2 
 + \xi_0 (p \cdot \hat{z})^2}\right],
 \label{eq:RS_gen}
\end{equation}
where $\hat{z}^\mu$ is the unit-vector along the rapidity coordinate and 
$g = 16$ is the number of gluonic degrees of freedom.
The parameters $\alpha_0$, $\Lambda_0$ and $\xi_0$ can be used 
to set the initial values $n_0$, $P_0$ and $\chi_0$,
as indicated in Eqs.~(11)--(13) of Ref.~\cite{Ambrus:2021sjg}.

In solving Eq.~\eqref{eq:boltz_bjorken}, it is convenient to take 
advantage of the azimuthal symmetry of the setup in both 
the coordinate and the momentum space. This allows only one point 
to be taken along the azimuthal direction $\varphi$, while $v_z$ 
can be discretized using the Gauss-Legendre quadrature of order 
$Q_\xi$ \cite{romatschke-prc-2011,ambrus-prc-2018}. Using the Gauss-Laguerre 
quadrature rules, $p$ is discretized using only $N_L = 2$ points, namely 
$p_1 = 2T_0$ and $p_2 = 6T_0$, where $T_0$ 
is taken as the initial temperature
\cite{ambrus-prc-2018}. Therefore, the 
total number of discrete momentum vectors employed for the simulations
presented in this subsection is $2Q_\xi$.
The expansions of $f^{\rm eq}$ and $f_{\rm RS}$ 
analogous to the one in Eq.~\eqref{eq:mj-feq-expansion-truncated} 
can be found in Refs.~\cite{ambrus-prc-2018} and \cite{ambrus-aip-2019},
respectively.

Labeling the discrete populations $f_{jk}$ with $1 \le j \le Q_\xi$ 
and $1 \le k \le N_L$, the 
derivatives of $f$ with respect to $v_z$ and $p$ can be computed 
via projection onto the Legendre and Laguerre polynomials, respectively
leading to linear relations. In the former case, we have 
\begin{equation}
 \left[\frac{\partial[v_z(1- v_z^2) f]}{\partial v_z}\right]_{jk} = 
 \sum_{j' = 1}^{Q_\xi} \mathcal{K}^P_{j,j'} f_{j',k},
\end{equation}
where the kernel matrix $\mathcal{K}^P_{j,j'}$ depending only on 
$Q_\xi$ can be precomputed. The explicit expression of its elements
can be found in Eq.~(3.54) of Ref.~\cite{ambrus-prc-2018}. 
For the derivative with respect to $p$, we can take advantage that 
$Q_L = 2$ is fixed and write
\begin{equation}
 \left[\frac{1}{p^2} \frac{\partial (fp^3)}{\partial p}\right]_{j,\frac{3}{2} \mp \frac{1}{2}} =
 \pm \frac{1}{2} f_{j1} \pm \frac{3}{2} f_{j2},
\end{equation}
where the upper and lower signs correspond to $k = 1$ and $2$, respectively.

We now compare the hydro and RTA solutions, focusing on the function $\chi$ 
given in Eq.~\eqref{eq:Bjorken_chi}. 
For further validation, we consider a comparison with 
the Boltzmann Approach to Multi-Parton Scattering (BAMPS), 
which is a particle-based stochastic method \cite{Xu:2004mz,Xu:2007jv}.
The initial state is prepared at vanishing chemical potential,
such that $n_0 \equiv n(\tau_0) = g T_0^3 / \pi^2$. 
Subsequently, the conservation of the particle number 
density implied by Eq.~\eqref{eq:Bjorken_eqs_n} enforces
$n(\tau) = n_0 \tau_0 / \tau$, leading to a non-trivial 
evolution of the chemical potential $\mu = T \ln(n \pi^2 / gT^3)$.
Enforcing a constant ratio between the shear viscosity $\eta = \frac{4}{5} \tau_R P$
 and the entropy density $s = (\epsilon + P - \mu n)/T$ fixes the relaxation time to
\begin{equation}
 \tau_R = \frac{5 \eta / s}{T} \left(1 + 
 \frac{3}{4} \ln \frac{\tau^{4/3} P}{\tau_0^{4/3} P_0}\right).
\end{equation}
In order to describe the evolution of $\chi$, it is convenient to employ the 
scaling variable $\tilde{w}$ defined via \cite{Blaizot:2021cdv}
\begin{equation}
 \tilde{w} = \frac{5 \tau}{4\pi \tau_R}.
 \label{eq:Bjorken_wtilde}
\end{equation}
In the case of a parton gas, for which $\mu = 0$ at all times, 
the above reduces to $\tilde{w} = \tau T / (4\pi\eta / s)$
\cite{Kamata:2020mka}. Within the parton gas model,
it was pointed out in Refs.~\cite{Heller:2015dha,Blaizot:2020gql,Blaizot:2021cdv} 
for the case of the hydro equations and in
Refs.~\cite{Heller:2016rtz,Blaizot:2017ucy,Strickland:2018ayk,Behtash:2019txb,Blaizot:2021cdv} for kinetic theory that
$\chi$ generally exhibits a decay from an arbitrary initial condition $\chi_0$ onto an attractor 
solution that bridges the free-streaming ($\tilde{w} = 0$) and 
the hydrodynamic ($\tilde{w} \rightarrow \infty$) fixed points. 
Figure~\ref{fig:bjorken} shows that a similar phenomenon occurs 
in the case of the ideal gas considered here. Here
we present the numerical results for the evolution of $\chi$ corresponding to two sets of $4 \times 4$ simulations,
one set for the RLBM results (solid lines) and another one for hydro (dashed lines). In addition, BAMPS results taken from Ref.~\cite{Ambrus:2021sjg} are shown with empty circles for 
a subset of curves.
The initial time and temperature 
are set to $\tau_0 = 0.2\ {\rm fm}$ and $T_0 = 0.5\ {\rm GeV}$, respectively, while $\eta / s$ takes 
the values $5$ (red) $1$ (blue) $0.2$ (green) and $0.05$ (purple), resulting in four 
different values of $\tilde{w}_0$. For each value of $\eta / s$, four 
initial values of $\chi_0$ are considered, namely $1$, $0.75$, $0.5$ and $0.25$.

The RTA and hydro attractor curves are shown with black solid and dashed lines, 
respectively. The analytical approximation for the RTA attractor derived in 
Ref.~\cite{Romatschke:2017vte} is also represented using a solid green line 
and is almost everywhere overlapped with our numerical solution.
The approach to the attractor can be clearly seen for both RLBM and 
hydro and most notably, these attractors differ when $\tilde{w} \lesssim 1$. In 
particular, it can be seen that the attractor solution for hydro gives $\chi < 0$ 
corresponding to an unphysical negative longitudinal pressure at small $\tilde{w}$. 
The agreement between hydro and RTA is restored when $\tilde{w} \gtrsim 1$, 
both at the level of the attractor solutions and of the dynamics of the approach 
to the attractor.
The BAMPS results are in excellent agreement with the 
RLBM solution throughout the entire flow regime.

\subsection{Anisotropic vortical flow}\label{subsec:elliptic_flow}

\begin{figure*}[tbh]
  \centering
  \includegraphics[width=0.99\textwidth]{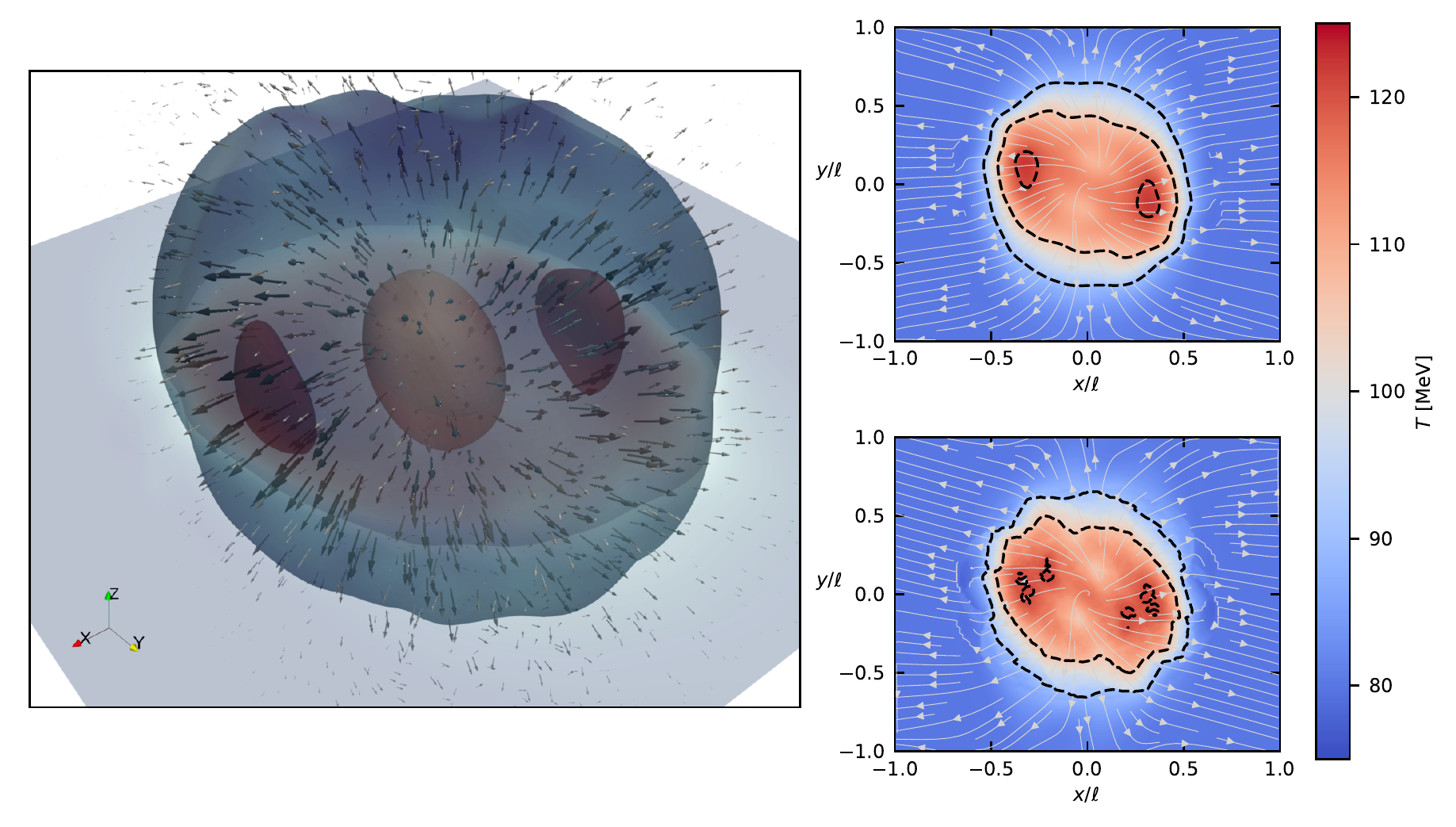}
  \caption{ Simulation of a vortical flow, using the initial conditions described in 
            Eq.~\ref{eq:rho_T_init} and Eq.~\ref{eq:U_init}, for a box of side $\ell = 20~\rm{fm}$ . 
            The three-dimensional figure in the left panel reports the temperature
            field after $t = 4~\rm{fm/c}$. We also provide a two-dimensional view
            on the $z=0$ plane at the same time step,  
            comparing the results provided by the model presented in this work (top-right),
            with the results obtained with a previous version of the model restricted to the simulation
            of hydrodynamic regimes (bottom-right). 
            The bottom-right panel shows that the interaction between QGP and the rarefied background
            gives origin to artifacts ("scars" in the $(x/\ell, y/\ell) \sim (\pm 0.5, 0.0)$ region
            and a general irregularity in the contour lines), not observed in the present model.
          } \label{fig:elliptic-flow} 
\end{figure*}

Recent measurements made by the STAR collaboration at the level of 
the decay products of the $\Lambda$ hyperons revealed that 
the quark-gluon plasma (QGP) formed during heavy ion collisions 
acquires a global polarization
\cite{STAR:2017ckg,STAR:2018gyt}. Possible
mechanisms leading to the polarization of the QGP constituents 
are the quantum chiral magnetic and chiral vortical effects \cite{Fukushima:2008xe,Kharzeev:2015znc} 
(see also Ref.~\cite{Ambrus:2020oiw} for an interplay between 
chiral and helical \cite{Ambrus:2019khr} vortical effects).
Taken together, 
these effects can explain the global polarization of the $\Lambda$
hyperons by means of a non-vanishing magnetic field or vorticity on 
the freezeout 
hypersurface. While the relevance of the chiral magnetic effect 
strongly depends on the lifetime of the magnetic field in the 
QGP fireball, vorticity is expected to be long-lived, decaying only 
due to dissipation caused by shear. 
Studies have estimated the vorticity to have a sizeable magnitude 
at freezeout \cite{Becattini:2015ska}. 
The polarization induced by vorticity can be estimated using the 
Wigner function formalism \cite{Becattini:2013fla}. Modeling the 
dynamics of vorticity using hydrodynamics gives an excellent 
match with the experimental data for the 
global polarization \cite{Karpenko:2016jyx} 
(i.e., along the total angular momentum vector $\bm{J}_{\rm sys}$).
The detailed structure of the local (differential) polarization along 
both the beam (or longitudinal) and the $\bm{J}_{\rm sys}$ directions
proves to be more challenging to reproduce, since non-equilibrium effects such 
as the coupling of spin with the thermal shear can make significant 
contributions to polarization 
\cite{Becattini:2021suc,Liu:2021uhn,Fu:2021pok,Becattini:2021iol}.

In this section, we show an example application of our new scheme aimed at 
simulating the dynamics of an initial vortex configuration in the 
more simplistic setup ignoring the longitudinal expansion 
(this was addressed in the frame of the Bjorken model in Sec.~\ref{subsec:attractor}).

We consider an ultra-relativistic gas in a cubic grid of side $20~\rm{fm}$, 
with open boundary conditions. 
Following previous works~\cite{friman-proc-2019, gabbana-ptrsa-2019}, we 
initialize the density and temperature fields with an asymmetric Gaussian shape:
\begin{align}\label{eq:rho_T_init}
        T &= T_b + T_0 \,g(x,y,z)  \; ,        \\ 
        n &= n_b + n_0 \,g(x,y,z)  \; , \notag \\ 
  g(x,y,z)&= \exp\left(-\frac{x^2}{2\sigma_x^2}-\frac{y^2}{2\sigma_y^2}-\frac{z^2}{2\sigma_z^2}\right) \; . \notag
\end{align}
Here $T_b = 80~\rm{MeV}$, $n_b = 10^{-3}~\rm{fm}^{-3}$ are background values for 
temperature and density, while $T_0=200~\rm{MeV}$, $n_0=4 \times 10^{-3}~\rm{fm}^{-3}$. 
We choose $\sigma_x = 1~\rm{fm}$, $\sigma_y = 2.6~\rm{fm}$ and $\sigma_z = 2~\rm{fm}$.
The initial velocity field is chosen as follows:
\begin{align}\label{eq:U_init}
        U_x/U_0 &= - \frac{y}{\sqrt{x^2+y^2}} \tanh{\left(\frac{\sqrt{x^2+y^2}}{r_0}\right)}\notag \; ,\\ 
        U_y/U_0 &= \phantom{+} \frac{x}{\sqrt{x^2+y^2}} \tanh{\left(\frac{\sqrt{x^2+y^2}}{r_0}\right)}       \; ,\\ 
        U_z/U_0 &= 0 \,\notag
\end{align}
with $r_0 = 6~\rm{fm}$.
We apply a cut-off radius $R = 3~\rm{fm}$ in the $z=0$ plane, outside 
of which the velocity field is set to zero.

The central ellipsoid represents the QGP formed in the collision between heavy nuclei. 
The highly compressed bulk of the system rotates and expands, cooling down in 
the process. In a later stage, the "fireball" further expands and cools down, so that
the system exits the hydrodynamic regime and enters a weakly interacting
rarefied regime known as "freeze-out".

The "freeze-out" regime is classified in terms of $\eta/s$, which in QGP is found
to reach the theoretical lower bound of $1/4\pi$~\cite{kovtun-prl-2005}. 

In order to characterize this effect, we have parametrized $\eta/s$ as a function
of the local temperature, using the following expression~\cite{zhang-jpg-2019, niemi-prl-2011}:  
\begin{align}
  \eta/s = 
  \begin{cases}
    0.681 - 0.0594\left(\frac{T}{T_R}\right) - 0.544 \left(\frac{T}{T_R}\right)^2  \; &T < T_R\\
    \frac{1}{4\pi}                                                                 \; &T \geq T_R
  \end{cases}
\end{align}
with $T_R = 175~\rm{MeV}$. In Fig.~\ref{fig:elliptic-flow}, we show the results of the simulation 
at $t = 4~\rm{fm/c}$, with the results obtained using a high order off-lattice
scheme ($N = 3$, $K = 15$) on the top right panel, and the results of the on-grid scheme
in the bottom left panel.

The comparison shows that the high order method allows to cure the artifacts clearly visible in 
the lower panel, in particular at the boundaries of the fireball where the fluid starts
to interact with the rarefied region.
%

\section{Performance data} \label{sec:performance}

In this section we give a short overview of the performances of the numerical model.
The present algorithm has been implemented for the benchmark described in 
Sec.~\ref{subsec:elliptic_flow} on a single V100-GPU machine using 
double-precision arithmetics and standard practice 
optimizations~\cite{calore-pc-2016,calore-cc-2016},
delivering about 60 MLUPS (Million Lattice UPdates per Second)
on a $128^3$ cubic grid with $N_{\rm pop} = 128$ discrete velocities. %
This means that the state of the system is advanced
over $30$ time steps in a second wall-clock GPU time.
For a computational box $10~\rm{fm}$ in side, this corresponds to a 
lattice spacing of about $0.1~\rm{fm}$ and a time-step of about $0.1~\rm{fm/c}$. 
The performance of our code is comparable to that of the GPU implementation of 
the ideal hydro code reported in Ref.~\cite{Gerhard:2012uf}, while giving access 
to off-equilibrium dynamics, such as dissipation.

\begin{figure}
  \centering
  \includegraphics[width=0.99\columnwidth]{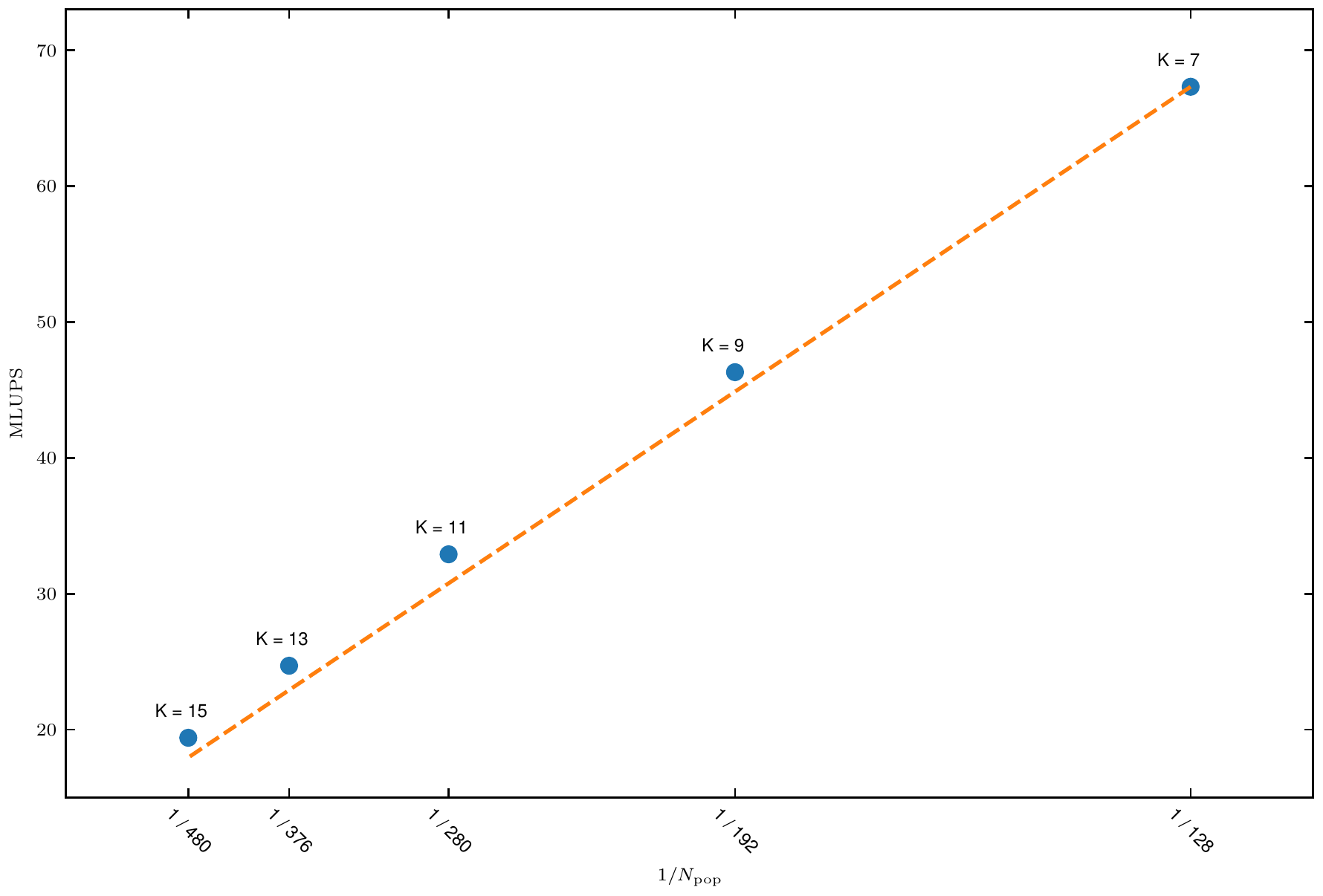}
  \caption{ Performance scaling with respect to the number of discrete velocities.
            Figures refer to quadratures for a ultra-relativistic gas with 
            a radial quadrature of degree $N = 3$ and an angular quadrature of degree $K$
            (see labels in the figure).
            The measured performances (blue dots) are measured in Million Lattice 
            UPdates per Second. The dashed line represents the linear 
            scaling with respect to $N_{\rm pop} = 128$. 
          } \label{fig:performance-cmp} 
\end{figure}

Notably, as shown in Fig.~\ref{fig:performance-cmp}, the 
performance scales linearly with the inverse number of components used in the 
discretization of the momentum space 
$1/N_{\rm pop}$, reaching down to about 20 MLUPS for the case $N_{\rm pop}=480$. 
This is an important result, as it shows that the code suffers no performance 
extra-penalty in going from the hydro to the quasi-ballistic regime.  

As a result, a simulation spanning one million time-steps, i.e. $10^5~\rm{fm/c}$, 
would complete in roughly $3 \times 10^4$ seconds, namely about half day.     
Since, as already observed, our method seamlessly describes both 
hydrodynamic and quasi-ballistic regimes, it can
efficiently simulate the long-term evolution of laboratory 
QGP well into the freeze-out regime and possibly beyond.

Indeed, with suitable coupling to Monte Carlo schemes, by sampling particle
position and momenta from the RLBM solution,~\cite{distaso-ptrsa-2016}, it may also be 
possible to describe the re-hadronization stage, in which quarks bind back into hadrons
\cite{fries-prl-2003, molnar-prl-2003, greco-prl-2003,Petersen:2008dd,Weil:2016zrk}. 
Compared to Monte-Carlo-based implementations such as BAMPS 
\cite{xu-prc-2007,bouras-prl-2009,Gallmeister:2018mcn,Ambrus:2021sjg}, which 
suffer from statistical noise, our scheme can be expected to be between $1$ and 
$2$ orders of magnitude faster.

On a mid-term perspective, one may realistically project the current data 
to large-scale massive parallel GPU architectures, such as the Nvidia A100 series.
For instance, recent work on multiphase non-relativistic fluids 
shows that classical Lattice Boltzmann schemes with 27 discrete populations can attain up to 
100 GLUPS on grids with several billion grid points, using large clusters with
hundreds Nvidia A100 GPUs~\cite{bonaccorso-cpc-2022, succi-cf-2019}.

Based on the linear dependence of the GPU performance on the inverse
number of populations, one can estimate about $20 \div 5$ GLUPS for the case of 
$100 \div 400$ discrete velocities. The same ballpark estimate is obtained by 
upscaling the current 60 MLUPS to 6 GLUPS on a hundred-GPUs cluster. 
This means several updates per second of grids with billion grid points, hence
enabling the direct simulation of QGP over three 
decades in space and twice as many in time, within a few days wall-clock time.

\section{Discussion}\label{sec:outlook}

Relativistic kinetic theory is ubiquitous to several fields of modern physics,
finding application both at large scales, in the realm of astrophysics, 
down to atomic scales (e.g., in the study of the electron properties of graphene)
 and further down to subnuclear scales, in the realm of quark-gluon plasmas. 
This motivates the quest for powerful and efficient computational methods, 
able to accurately study fluid dynamics in the relativistic regime as well as 
the transition to beyond hydrodynamics, in principle all the way down to ballistic regimes.

In this work we have introduced a lattice kinetic 
scheme which extends the range of applicability 
of RLBM to a wider range of kinetic parameters,
allowing for the simulation of relativistic gases 
of massive particles in rarefied conditions.

The present scheme builds on high order quadrature rules
developed to separately discretize the radial and the angular 
coordinates in the momentum space.
These quadratures are in general not compatible with
a Cartesian grid, therefore introducing the need for
an interpolation scheme in the streaming step.
We show that by increasing the degree of the radial and angular
quadrature it is possible to tune the accuracy of the numerical scheme
to the given kinetic parameters. 

By analyzing shock waves in a quark-gluon plasma we have shown 
that in order to achieve good accuracy in rarefied conditions,
the order of the radial quadrature must increase at increasing
values of the relativistic coldness.
This means that the non-relativistic regime is more demanding
than the relativistic one, which is in line with the fact that non-relativistic
particles move in a broader range of speeds as compared to the
relativistic ones. 

From a computational point of view, RLBM retains the 
advantages of standard Lattice Boltzmann schemes, making
it an ideal candidate for efficient implementations on
massively parallel architectures.

We have evaluated the performances of a GPU implementation of the method,
showing that the computational cost grows linearly with
the number of discrete components employed in
the momentum space discretization.

This paves the way to the systematic study of heavy-ion collisions observables 
such as the $p_T$ dependence of the flow harmonics $v_n(p_T)$ 
\cite{Romatschke:2018wgi} or hadron polarization \cite{Karpenko:2016jyx} within 
kinetic theory.
Extensions of the current scheme to the case of non-ideal fluids can be performed
along the lines discussed in Ref.~\cite{romatschke-prd-2012}, allowing phenomena 
related to the QCD phase transition to be explored through large-scale 
simulations \cite{STAR:2013gus,Habich:2014tpa,Nahrgang:2018afz}.

While a fair comparison between RLBM and Monte Carlo approaches
is somehow ill-posed, since the latter can handle non-equilibrium
effects in full, for systems where the relaxation-time approximation applies,
RLBM can be expected to offer one or two orders of magnitude speedup over
Monte Carlo methods. 

To conclude, the present results lay the ground to the computationally 
efficient large-scale simulations of beyond-hydrodynamic 
regimes in the framework of QGP experiments. They may also find profitable use
in the study of quasi-ballistic electron flows in graphene and possibly also  
for relativistic flows of astrophysical interest. In addition, the implementation
of the Boltzmann-Vlasov equation for resistive relativistic magnetohydrodynamics
\cite{Denicol:2019iyh,Bacchini:2022myw} is straightforward via the addition of
the electromagnetic forcing term, and this in turn might unlock applications in
the realm of Plasma Wakefield Acceleration \cite{parise-pp-2022}.

\section{Methods}\label{sec:model-description}

In this section we provide full details on the definition of the Relativistic 
Lattice Boltzmann Method, with particular emphasis posed on the momentum space 
discretisation, which is crucial in order to support the simulation of dynamics
at large values of the Knudsen number.
We start by introducing the notation and with a brief introduction to the main
elements of relativistic kinetic theory.

\subsection{Relativistic Kinetic Theory}\label{subsec:relativistic-kinetic-theory}

We consider a gas of particles with mass $m$ in a $(3+1)$ Minkowski space-time,
with metric $\eta^{\alpha\beta}={\rm diag}(+,-,-,-)$. We adopt Einstein's 
summation convention, with Greek indices running from $0$ to $3$, and latin ones
from $1$ to $3$, respectively. We also use natural units: 
$c = k_{\rm B} = \hbar = 1$. 

Our starting point in the development of the model is the relativistic Boltzmann
equation, in the single-relaxation time approximation of Anderson and 
Witting~\cite{anderson-witting-ph-1974b,anderson-witting-ph-1974a} 
\begin{align}\label{eq:boltz_eq}
  p^{\alpha}\partial_\alpha f = 
  - \frac{U^\alpha p_\alpha}{\tau} \left( f - f^{\rm eq} \right) \quad .
\end{align}
This equation describes the evolution of the particle distribution function 
$f(x^{\alpha}=(t,\mathbf{x}), p^{\alpha}=(p^{0}, \mathbf{p}))$, accounting for
the number of particles per unit volume in the 6-dimensional single-particle 
phase space $\diff^3 x \diff^3 p$.

$U^\alpha$ is the macroscopic fluid velocity, $\tau$ is the (proper)-relaxation 
time and $f^{\rm eq}$ is the equilibrium distribution function:
\begin{align}\label{eq:feq_dist}
  f^{\rm eq}(p^{\alpha}, U^{\alpha}, T) = 
   \frac{ (2\pi)^{-3} g}{\exp{\left(\frac{p^\alpha U_\alpha - \mu}{T}\right)} + \varepsilon} 
   \quad , 
\end{align}
where $\mu$ is the chemical potential, $g$ is the number of degrees of freedom 
per constituent, and $\varepsilon$ is a parameter selecting between the 
Maxwell-J{\"u}ttner  ($\varepsilon=0$), 
Fermi-Dirac ($\varepsilon=1$) and Bose-Einstein ($\varepsilon=-1$) distributions.
From now on, we will be considering the Maxwell-J{\"u}ttner statistics 
($\varepsilon=0$).

The chemical potential $\mu$ can be expressed in terms of the particle number 
density. For the Maxwell-J\"uttner statistics, we have
\begin{align}
    \frac{g}{(2\pi)^3} e^{\mu/T} = \frac{n}{4\pi T^3 \zeta^{2} K_{2}(\zeta )} \quad ,
\end{align}
with $n$ the particle number density, and $K_\nu$ the modified Bessel function 
of second kind of index $\nu$.

The hydrodynamic fields are related to the lower order moments of the 
distribution function, in particular the first and second order moments, namely
the particle flow $N^{\alpha}$ and energy momentum tensor $T^{\alpha\beta}$, 
\begin{align}
  N^{\alpha} = \int p^\alpha f \frac{\diff^3 p}{p^0}                   \quad , \label{eq:moments} \\
  T^{\alpha\beta} = \int p^\alpha p^\beta f \frac{\diff^3 p}{p^0}      \quad . \label{eq:moments2}
\end{align}
The moments of the distribution can be put in relation to the macroscopic fields
via the Landau-Lifschitz~\cite{landau-book-1987} decomposition:
\begin{align} \label{eq:landau-decomposition}
  N^{\alpha}    &= n U^\alpha - \frac{n}{P + \epsilon} q^\alpha  
  \quad ,  \\
  T^{\alpha\beta} &= (P + \epsilon + \varpi ) U^{\alpha}U^{\beta}   
           - (P + \varpi) \eta^{\alpha\beta} 
           + \pi^{<\alpha\beta>}                                     \quad ;
\end{align}
where $P (\varpi)$ is the hydrostatic (dynamic) pressure, $q^\alpha$ the heat
flux, $\epsilon$ the energy density and $\pi^{<\alpha\beta>}$ the pressure 
deviator. By computing the integrals in Eq.~\ref{eq:moments} and~\ref{eq:moments2}
at equilibrium, and matching them with the known expressions of the equilibrium
moments $N^{\alpha}_{\mathrm{eq}}$ and $T^{\alpha\beta}_{\mathrm{eq}}$, one finds
the following ideal Equation of State (EoS):
\begin{align} \label{eq:eos}
  \epsilon = P \left(\zeta \frac{ K_{3}(\zeta)}{K_{2}(\zeta)} - 1 \right)
  \quad , \quad
  P = n T   \quad ,
\end{align}
This reduces to the familiar expressions $\epsilon = 3 P$ in the 
ultrarelativisitc case ($\zeta \to 0$) and $P = (3/2) nT$ in the non-relativistic
one ($\zeta \to \infty$), respectively.

\subsection{Relativistic Lattice Boltzmann Method}\label{subsec:RLBM}

We start, after an adimensionalization of variables, by considering a $N$-truncated 
expansion of the Maxwell-J{\"u}ttner distribution (Eq.~\ref{eq:feq_dist}) onto a
tensorial basis of rank-$k$, $\mathbf{J}^{(k)}(p^\mu)$:
\begin{equation}\label{eq:mj-feq-expansion-truncated}
  f^{\rm eq}(p^{\mu}, U^{\mu}, T) 
  = 
  \omega(p^0) \sum_{k = 0}^{N} \mathbf{a}^{(k)}( U^{\mu}, T) \cdot \mathbf{J}^{(k)} ( p^{\mu} )  
  \quad ,
\end{equation}
where ``$\cdot$'' represents full tensor contraction.
These tensors $\mathbf{J}^{(k)}$ are built as orthogonal polynomials in the 
variable $p^\mu$ with respect to a weighting function $\omega(p^{0})$, by using
a standard Gram-Schmidt procedure, and can be shown to satisfy the following 
orthonormality condition:
\begin{equation}\label{eq:orthonormal-condition}
 \int \omega(p^{0}) \, \mathbf{J}^{(l)}_\alpha(p^{\mu}) \, \mathbf{J}^{(k)}_\beta ( p^{\mu}) \frac{\mathrm{d}^{3} p}{p^{0}}
 = 
 \delta^{lk}\delta_{\alpha \beta} 
 \quad,
\end{equation}
where $\alpha=\{\alpha_1,\dots, \alpha_l\}$ and $\beta=\{\beta_1,\dots, \beta_k\}$
are collective tensorial indices introduced for notational conciseness. A 
detailed discussion and derivation of this set of orthogonal polynomials can be
found in \cite{gabbana-pr-2020}. 
The expansion coefficients in Eq.~\ref{eq:mj-feq-expansion-truncated} are 
defined as
\begin{align}\label{eq:mj-projection-coefficients}
  \mathbf{a}^{(k)}( U^{\mu}, T) 
  = 
  \int f^{\rm eq}( p^{\mu}, U^{\mu}, T) \, \mathbf{J}^{(k)}( p^{\mu} ) \frac{\diff^3 p}{p^0} \quad.
\end{align}

The choice of the weight function $\omega(p^0)$ is instrumental: by taking it as
the equilibrium distribution in the rest frame, it is possible to establish a 
direct link between each coefficient $\mathbf{a}^{(k)}$ and the corresponding 
$k-th$ moment of the distribution function.

Next, we define a quadrature rule satisfying the requirement of preserving all the moments of the 
distribution up to order $N$.
The quadrature is obtained as product of Gaussian quadratures: we consider a radial quadrature of 
degree $N$, consisting of $(N+1)$ discrete components, and an angular quadrature of degree $K$,
consisting of $N_K$ discrete components (see Sec.~\ref{subsec:discretization} for full details).
This results in a set of $N_{pop} = N_K (N+1)$ discrete momenta $\{ p^\mu_i \} $ and corresponding 
weights $\{ w_i \}$, which allows to define the discretized version of the equilibrium distribution
\begin{equation}\label{eq:discrete-mj-feq-expansion-truncated}
  f^{\rm eq}_i = f^{\rm eq}(p^{\mu}_i, U^{\mu}, T) 
  = 
  w_i \sum_{k = 0}^{N} \mathbf{a}^{(k)}( U^{\mu}, T) \cdot \mathbf{J}^{(k)} ( p^{\mu}_i )  \quad .
\end{equation}
By construction this recovers all the moments of the distribution function in 
the continuum up to order $N$, and it follows that integrals in 
Eq.~\ref{eq:moments} and ~\ref{eq:moments2} can be computed \textit{exactly} 
(i.e. equality holds) via discrete sums:
\begin{align}\label{eq:discrete_sum_moments}
  N^\alpha = \sum_i^{N_{\rm pop}} p^\alpha_i f_i      
  \quad , \quad  
  T^{\alpha\beta} = \sum_i^{N_{\rm pop}} p^\alpha_i p^\beta_i f_i \quad ,
\end{align}
where $f_i = f_i (\bm{x}, t) = f (p^{\mu}_i, \bm{x}, t)$. 
By combining the quadrature-based discretization of the momentum space with a 
forward-Euler discretization in time with time-step $\Delta t$, it is possible 
to derive the discrete relativistic Lattice Boltzmann equation:
\begin{equation}\label{eq:discrete-rbe}
  f_i(\bm{x} + \bm{v}_{i} \Delta t, t + \Delta t) 
  = 
  f_i(\bm{x}, t) + \Delta t~ \frac{p_i^{\alpha} U_{\alpha}}{p^0_i \tau} 
                   (f_i^{\rm eq} - f_i(\bm{x}, t) ) \; ,
\end{equation}

where $\bm{v}_{i} = \bm{p}_i / p^0_i $. 

We conclude this section with a  quick summary of the algorithmic procedure 
needed to advance Eq.~\ref{eq:discrete-rbe} over a single time step, based on 
the stream\&collide paradigm.

Starting from a suitable initialization $f_i(t=0)$, at each time step the 
discrete populations freely stream to the corresponding lattices sites:
\begin{align}\label{eq:streaming}
  f_i^*(\bm{x}, t) = f_i(\bm{x} - \bm{v}_{i} \Delta t, t) \quad ,
\end{align}
This moves information from each lattice point at a distance 
$\Delta_i \bm{x} = \bm{v}_{i} \Delta t$.
Clearly, an interpolation is required whenever $\bm{x} - \Delta_i \bm{x}$ does 
not fall on a grid point, in order to infer the values of the populations at 
the nodes of the actual Cartesian grid, based on their off-lattice values. In 
this work we adopt a simple trilinear interpolation scheme:
\begin{equation}\label{eq:interpolation}
\begin{split}
  f_i (\bm{x} - \bm{v}^{i} \Delta t, t) = &\frac{1}{\Delta x \, \Delta y\, \Delta z} \times \Big\{   \\
     & f_i(\bm{x} -\bm{r}_x -\bm{r}_y - \bm{r}_z, t) 
     \big( \phantom{\Delta x - {}} \Delta t \big| v^i_x \big| \big) 
     \big( \phantom{\Delta y - {}} \Delta t \big| v^i_y \big| \big) 
     \big( \phantom{\Delta z - {}} \Delta t \big| v^i_z \big| \big)   \,+  \\  
     & f_i(\bm{x}   \phantom{ {} -\bm{r}_x} -\bm{r}_y -\bm{r}_z, t) 
     \big(          \Delta x -     \Delta t \big| v^i_x \big| \big) 
     \big( \phantom{\Delta y - {}} \Delta t \big| v^i_y \big| \big)     
     \big( \phantom{\Delta z - {}} \Delta t \big| v^i_z \big| \big)   \,+  \\  
     & f_i(\bm{x} -\bm{r}_x   \phantom{ {} -\bm{r}_y} -\bm{r}_z, t) 
     \big( \phantom{\Delta x - {}} \Delta t \big| v^i_x \big| \big) 
     \big(          \Delta y -     \Delta t \big| v^i_y \big| \big)      
     \big( \phantom{\Delta z - {}} \Delta t \big| v^i_z \big| \big)   \,+  \\  
     & f_i(\bm{x} -\bm{r}_x -\bm{r}_y   \phantom{ {} -\bm{r}_z}, t) 
     \big( \phantom{\Delta x - {}} \Delta t \big| v^i_x \big| \big) 
     \big( \phantom{\Delta y - {}} \Delta t \big| v^i_y \big| \big)
     \big(          \Delta z -     \Delta t \big| v^i_z \big| \big)   \,+  \\  
     & f_i(\bm{x}  -\bm{r}_x   \phantom{ {} -\bm{r}_y}   \phantom{ {} -\bm{r}_y}, t) 
     \big( \phantom{\Delta x - {}} \Delta t \big| v^i_x \big| \big)
     \big(          \Delta y -     \Delta t \big| v^i_y \big| \big)   
     \big(          \Delta z -     \Delta t \big| v^i_z \big| \big)   \,+  \\  
     & f_i(\bm{x}   \phantom{ {} -\bm{r}_x}  -\bm{r}_y   \phantom{ {} -\bm{r}_y}, t) 
     \big(          \Delta x -     \Delta t \big| v^i_x \big| \big) 
     \big( \phantom{\Delta y - {}} \Delta t \big| v^i_y \big| \big)     
     \big(          \Delta z -     \Delta t \big| v^i_z \big| \big)   \,+  \\  
     & f_i(\bm{x}   \phantom{ {} -\bm{r}_y}   \phantom{ {} -\bm{r}_y} -\bm{r}_z, t) 
     \big(          \Delta x -     \Delta t \big| v^i_x \big| \big)  
     \big(          \Delta y -     \Delta t \big| v^i_y \big| \big)     
     \big( \phantom{\Delta z - {}} \Delta t \big| v^i_z \big| \big)   \,+  \\  
     & f_i(\bm{x}   \phantom{ {} -\bm{r}_x} \phantom{ {} -\bm{r}_y}   \phantom{ {} -\bm{r}_z}, t) 
     \big(          \Delta x -  \Delta t \big| v^i_x \big| \big) 
     \big(          \Delta y -  \Delta t \big| v^i_y \big| \big)
     \big(          \Delta z -  \Delta t \big| v^i_z \big| \big) \quad\Big\} 
\end{split}
\end{equation}
with 
\begin{equation}
\begin{cases}
  \bm{r}_x = \mathrm{sgn}(v^i_x) \,(\Delta x) \, \hat{\mathbf{x}}  \\
  \bm{r}_y = \mathrm{sgn}(v^i_y) \,(\Delta y) \, \hat{\mathbf{y}}  \\
  \bm{r}_z = \mathrm{sgn}(v^i_z) \,(\Delta z) \, \hat{\mathbf{z}}  \quad,
\end{cases}
\end{equation}
$(v^i_x,v^i_y,v^i_z)$ being the components of the velocity vectors in the 
stencil.

Next, the first and second order moments are computed using Eq.~\ref{eq:discrete_sum_moments}. 

Thermodynamic quantities can be recovered from Eq.~\ref{eq:eos}, after
solving the following eigenvalue problem:
\begin{align}\label{eq:eigvalue_problem}
  \epsilon U^{\alpha} &= T^{\alpha \beta} U_{\beta}         \quad , \\
    n &= U^\alpha N_\alpha                           \notag \quad .
\end{align}

At this stage, it is possible to compute the new local equilibrium distribution
(Eq.~\ref{eq:discrete-mj-feq-expansion-truncated}), which is needed 
to apply the collisional operator:
\begin{align}\label{eq:collision}
  f_i(\bm{x}, t + \Delta t) = f_i^*(\bm{x}, t) 
                            + \Delta t~ \frac{p_i^{\alpha} U_{\alpha}}{p^0_i \tau} 
                            (f_i^{\rm eq} - f_i^*(\bm{x}, t)) \quad .
\end{align}

\subsection{Momentum space discretization} \label{subsec:discretization}

In this section we present a detailed discussion of the momentum space 
discretization.
We make use of off-lattice quadratures,  which are developed as product of 
Gaussian quadratures~\cite{bazzanini-jocs-2021, ambrus-prc-2018}, offering the 
possibility of handling more complex equilibrium distribution functions and, in 
turn, extending the applicability of the method to regimes beyond hydrodynamics.

We define a quadrature of order $N$ as a quadrature having the property of 
preserving exactly (i.e., equality holds when integrals are calculated with 
discrete summations) the first $N$ moments of the particle distribution. 
Formally, this can be expressed by requiring that all the integrals in 
  the form:
\begin{align}\label{eq:quad_integrals}
  I^{\alpha_1 \dots \alpha_k} = \int \omega(p_0) \, p^{\alpha_1} \cdots p^{\alpha_k} \frac{\diff^3 p}{p_0} \quad ,
\end{align}
must be exactly computed by the quadrature $\forall k \leq 2N$. 

As already stated, the weight function $\omega(p^0)$ is proportional to the 
equilibrium distribution function computed in the rest frame ($U^\alpha=(c,0,0,0)$) 
\begin{align}\label{eq:feq_dist}
  \omega(p^0) = 
   C \exp{\left(-\frac{p^0}{T}\right)}  
   \quad , 
\end{align}
with $C$ a factor such that $\omega(p^0)$ is normalized to unity.

By introducing the following change
of variables 
\begin{align}\label{eq:y_coord_change}
  \begin{cases}
    p^0 &= y + m                                 \\
    p^x &= \sqrt{y(y+2m)} \sin\theta\cos\varphi  \\
    p^y &= \sqrt{y(y+2m)} \sin\theta\sin\varphi  \\
    p^z &= \sqrt{y(y+2m)} \cos\theta            
  \end{cases} \quad ,
\end{align}
Eq.~\ref{eq:quad_integrals} can be split into two parts
\begin{equation}
  I^{\alpha_1\dots\alpha_k} = I_R \times I_\Omega  \quad ,
\end{equation}
respectively the angular part  $I_\Omega$
\begin{equation}\label{eq:angular_part}
  I_\Omega = \int (\sin\theta\cos\varphi)^{k_x}(\sin\theta\sin\varphi)^{k_y}(\cos\theta)^{k_z}\diff \Omega  \; ,
\end{equation}
and the radial part $I_R$
\begin{equation}\label{eq:radial_part}
  I_R = \int_{0}^{+
  \infty} W(y) \; Q(y) \; \diff y ,
\end{equation}
with 
\begin{align}
    k  &= k_0+k_x+k_y+k_z = k_0 + K                                         \\
  W(y) &=
  \sqrt{y(y+2m)} \; 
  \omega(y+m)
  \label{eq:weight} \\
  Q(y) &= (y+m)^{k_0} (y^2+2my)^{\frac{K}{2}}           \label{eq:poly}    
\end{align}
and all $k_0$, $k_x$, $k_y$ and $k_z$ accounting for the number of occurrences 
of the various degrees of freedom in $I^{\alpha_1\dots\alpha_k}$. 

\subsubsection{Radial Discretization}

We focus now on the discretization of the radial integrals. 
We consider $I_R$ with $K$ an even number, since by symmetry the angular 
integral $I_\Omega$ cancels out for odd values of $K$.

From Eq.~\ref{eq:poly} we observe that in this case $Q(y)$ is a polynomial of 
degree $k$, and therefore it is possible to establish a Gauss-like quadrature 
rule to perform an exact integration of $I_R$. To this aim we consider the 
following polynomial basis:
\begin{align}
     P_0 = \mathbf{J}^{(0)}      \;,
     P_1 = \mathbf{J}^{(1)}_{0}  \;,
  \,\dots\,                   ,
  P_{2N} &= \mathbf{J}^{(2N)}_{0\dots0}
\end{align}
that constitutes an orthogonal basis with respect to the weight $W(y)$ 
defined in Eq.~\ref{eq:weight}; here the polynomials $\mathbf{J}^{(k)}$ 
are the ones introduced before in Eq.~\ref{eq:mj-feq-expansion-truncated}, 
and are taken with all indices equal to zero. 
By referring to the theory of Gaussian Quadratures~\cite{abramowitz-book-1965}, 
one can derive the $N$-th order radial quadrature rule in the following way:
\begin{align}
  \text{abscissae }     y_i:&\quad  \text{roots of } P_{N+1} (y) \quad , \\
  \text{weights } w_i^{(y)}:&\quad  \int_0^{+\infty} \frac{W(y) P_{N+1} (y) }{(y-y_i)\,P'_{N+1}(y_i)} \mathrm{d}y\quad .
\end{align}
The corresponding values for the discrete energy, the absolute value of the momentum and 
velocity can be recovered from the discrete coordinate $y_i$ through Eq.~\ref{eq:y_coord_change}. 

For the special case $m = 0$ our procedure coincides with the
generalized Gauss-Laguerre quadrature rule. 

\subsubsection{Angular Discretization}

Let us now turn to the discretization of the angular part; notice that the angular 
integral is independent on the mass of the particles. One has
\begin{align}\label{eq:angular_int_3d}
  I_{\Omega} = \int (\sin\theta \cos\varphi)^{k_x} (\sin \theta\sin\varphi)^{k_y} (\cos \theta)^{k_z} \diff \Omega \quad .
\end{align}
The integrand can be recasted into a sum of spherical harmonics $Y_{\ell}^{m}(\theta, \varphi)$ of 
maximum degree $K$. Therefore any spherical quadrature that integrates exactly all spherical 
harmonics up to order $\ell=K$ is a proper candidate for our goal. We therefore shift the problem to
the exact discrete computation of 
\begin{align}\label{eq:spherical_harmonics_quad}
  \int  Y_{\ell}^{m}(\theta, \varphi) \, \mathrm{d} \Omega=\sum_{q=1}^{N_{\mathrm{pop}}} w_{q} Y_{\ell}^{m}\left(\theta_{q}, \varphi_{q}\right), \quad \forall  \ell \leq K \quad .
\end{align}

Several different spherical quadrature rules are available in the literature 
(see e.g.~\cite{weih-mnras-2020} for a few examples).
In this work we adopt spherical design quadratures~\cite{delsarte-gd-1977}, and in particular we use
the sets of stencils defined in~\cite{womersley-book-2018}.

\subsubsection{Decoupling of the radial and angular quadratures}

With the procedures described in the previous sections, the nodes and weights of the whole 
stencil are expressed as 
\begin{align}
  \begin{cases}
    p^0_{ij} &= y_i + m                                 \\
    p^x_{ij} &= \sqrt{y_i(y_i+2m)} \sin\theta_{j} \cos \varphi_{j} \\
    p^y_{ij} &= \sqrt{y_i(y_i+2m)} \sin\theta_{j} \sin \varphi_{j} \\
    p^z_{ij} &= \sqrt{y_i(y_i+2m)} \cos\theta_{j} 
  \end{cases}\quad,
\end{align}
\begin{align}
  w_{ij} &= w^{(y)}_i w^{(\theta,\varphi)}_j 
  \quad , \quad 
  \begin{aligned}
    i &= 1, \dots, N+1 \\
    j &= 1, \dots, N_K 
  \end{aligned} \quad .
\end{align}
The (minimum) number of discrete components required to implement the quadrature is then 
$N_{\rm pop} = N_K (N+1)$. 
When working in the hydrodynamic regime one is generally interested in defining the quadrature
with the minimal number of discrete components, in order to minimize the computational cost 
of the numerical method.

On the other hand, when moving to regimes characterized by high values of the Knudsen number, 
stencils with more than the minimum amount of required discrete velocities are needed, since, as the
gas becomes more and more rarefied, even small errors in the velocities space become increasingly 
detrimental to the numerical solution. 

One way to achieve better solutions is therefore to increase the number of discrete velocities per 
energy shell, which however comes at an increased computational cost. 
Another possible action that enhances the solution is the decoupling of the radial and angular 
abscissae; indeed, once we have accepted to work off-lattice and once we have granted the required 
isotropy level for recovering the requested moments of the distributions, the restriction of using 
the same angular stencils for each energy shell $p_i^0$ can be relaxed. In this way, one can enhance
 the isotropy of the stencil without having to increase the whole quadrature order.

In $(2+1)$ dimensions this is easily achieved by rotating the sub-stencils related to different 
energy shells each with a different angle, in such a way that the discrete velocities cover the 
velocity space in the most homogeneous possible way. Further details can be found 
in~\cite{bazzanini-jocs-2021} for the (2+1) ultra-relativistic case. 

In $(3+1)$ dimensions the decoupling process is not trivial anymore, since we have a relative 
freedom in the specification of the rotations between the sub-stencils.  In fact, having 
considered an initial velocity set, derived using one of the spherical design 
quadrature exposed above, then one has, for a radial quadrature of order $N$, $N + 1$ 
overlapped shells of vectors $G_i$ belonging to the set $G = \bigcup_i^{N+1} G_i$.                        

Then one has to determine the set of angles $\{\alpha_i, \gamma_i\}$, with $i = 1 \dots N+1$, that 
defines the rotation matrix
\begin{equation}
R(\alpha_i,\gamma_i) = 
  \begin{pmatrix}
  \cos\alpha_i \cos\gamma_i & -\sin\alpha_i & \cos\alpha_i \sin\gamma_i \\
  \sin\alpha_i \cos\gamma_i &  \cos\alpha_i & \sin\alpha_i \sin\gamma_i \\ 
             - \sin\gamma_i &             0 &              \cos\gamma_i 
  \end{pmatrix} \quad .
\end{equation}
The new stencil $G'$ is then defined as $G' = \bigcup_i^{N+1} R(\alpha_i, \gamma_i) \cdot G_i$. 

There are several approaches with which one can find the different rotation matrices 
$R(\alpha_i, \gamma_i)$. Here, we adopt the following:

\begin{itemize}
  \item Once a radial discretization order $N$ is set, one obtains $N+1$ energy shells, 
  and consequently $N+1$ velocity subsets $G_i$, $i=1,\dots,N+1$.
  \item Depending on the value of $N$ one adopts the following strategies: 
  \begin{itemize}
    \item When $N + 1 = (4, 6, 8, 12, 20)$, we identify Platonic Solids with $N+1$
          vertexes. Then, the rotation matrices $R(\alpha_i, \gamma_i)$ are the ones that map 
          one vertex of the solid to its other vertexes. 
    \item Instead, for generic values of $N$ the $R(\alpha_i, \gamma_i)$ are determined by solving the 
          Thomson problem~\cite{thompson-jofm-1986}, that is related to the minimization of 
          electrostatic energy of electrons constrained on the sphere. Indeed, by treating discrete velocities 
          as electrons, one can determine the $R(\alpha_i, \gamma_i)$ matrices by iteratively 
          joining the substencils $G_i$ and solving the associated Thomson problem for $\alpha_i$ and 
          $\gamma_i$.
  \end{itemize}
\end{itemize}

\section*{Acknowledgment}

DS has been supported by the European Union's Horizon 2020 research and
innovation programme under the Marie Sklodowska-Curie grant agreement No. 765048.
SS acknowledges funding from the European Research Council under the European
Union's Horizon 2020 framework programme (No. P/2014-2020)/ERC Grant Agreement 
No. 739964 (COPMAT). VEA gratefully acknowledges the support of the Alexander 
von Humboldt Foundation through a Research Fellowship for postdoctoral 
researchers.
All numerical work has been performed on the COKA computing cluster at 
Universit\`a di Ferrara. 

\section*{Data Availability}

The data that support the findings of this study are available from the corresponding
author upon reasonable request.

\section*{Code Availability}

The code, along with examples for running the Riemann problem, data and scripts 
for reproducing Fig.~1, 2 and 4, have been deposited to Code Ocean~\cite{gabbana-code-2022}

\newpage
\bibliography{biblio}

\end{document}